\documentclass[aps,pre,longbibliography,notitlepage,floatfix,nofootinbib]{revtex4-1}

\usepackage{bm,amsmath,amssymb,graphicx,stmaryrd,float,subcaption,upgreek,sidecap,dsfont,mathtools,gensymb}
\usepackage[abs]{overpic}
\newcommand{\uvc}[1]{\bm{\mathrm{\hat #1}}} 
\usepackage{comment,footnote}
\usepackage[format=plain,labelfont={bf,small},textfont=small,justification=raggedright,singlelinecheck=false]{caption}

\newcommand{\bX}{{\bm x}}
\newcommand{\shift}[1]{#1^*}
\newcommand{\shiftvar}[1]{\bar{#1}}

\newcommand{\bOmega}{{\bm \Omega}}
\newcommand{\bn}{{\bm n}}
\newcommand{\bmo}{{\bm m}}
\newcommand{\bP}{{\bf P}}

\newcommand{\bC}{{\bf C}}

\newcommand{\bM}{{\bf M}}

\newcommand{\bR}{{\bf R}}
\newcommand{\bJ}{{\bf J}}
\newcommand{\bQ}{{\bf Q}}

\begin{document}
	\title{On the planar \emph{elastica}, stress, and material stress}
	\author{H. Singh}
	\email{harmeet@vt.edu}
	\affiliation{Department of Biomedical Engineering and Mechanics, Virginia Polytechnic Institute and State University, Blacksburg, VA 24061, U.S.A.}
	\author{J. A. Hanna}
	\email{hannaj@vt.edu}
	\affiliation{Department of Biomedical Engineering and Mechanics, Department of Physics, Center for Soft Matter and Biological Physics, Virginia Polytechnic Institute and State University, Blacksburg, VA 24061, U.S.A.}
	\date{\today}

\begin{abstract}
We revisit the classical problem of the planar Euler \emph{elastica} with applied forces and moments, and present a classification of the shapes in terms of tangentially conserved quantities associated with spatial and material symmetries.  We compare commonly used director, variational, and dynamical systems representations, and present several illustrative physical examples.  
We remark that an approach that employs only the shape equation for the tangential angle obscures physical information about the tension in the body.
\end{abstract}

\maketitle

\section{Introduction}

The planar Euler \emph{elastica} \cite{Oldfather33} has the simplest curvature-dependent energy for a one-dimensional continuum.  As a result, it appears in idealized models of a variety of physical systems, from structural cables and thin beams to biological macromolecules \cite{Mladenov17}.
This breadth of application has given rise to numerous derivations of the governing equations across the scientific and engineering literature, often with specific boundary conditions imposed \emph{a priori}.  Some of these derivations are quite lengthy, and may give the student of elastic structures the impression that the system is far more complicated than it really is.  The possibility of end moments is often ignored in expository treatments, and we know of no place where any two of the three most common approaches to the \emph{elastica} are presented together, or the connections between them laid out.  We suspect that much of this knowledge is well known among practitioners of mechanics, but feel that it is worth collecting in one place, alongside new results and perspectives.

In this note, we attempt to clarify some of these folkloric issues, as well as present a reclassification of the \emph{elastica} curves in terms of quantities conserved under translations in material and physical space.  The curves represent thin elastic bodies with material symmetry along their long axes, embedded in a symmetric Euclidean background.  The associated material momentum and conventional (spatial) linear momentum balances express the conservation of material and conventional stresses.  We will refer to the latter simply as the stress in what follows.\footnote{Material forces and closely related quantities, which in our one-dimensional system have just one component, appear under many names in the literature, including Eshelbian force, quasimomentum, pseudomomentum, (Kelvin) impulse, and configurational force \cite{Rogula77, Herrmann81, Benjamin84, GurevichThellung90, Nelson91, Maugin95, KienzlerHerrmann00, Gurtin00, Yavari06, OReilly2017}.  In the present case, one can derive everything from consideration of conventional force balance and its projection onto the tangents of the body, but the concept of material force is useful as a descriptive term, and also corresponds to an important symmetry of the Lagrangian.  We generally prefer the term ``spatial'' to ``conventional'', but try to avoid it in this note because of its alternative meaning as ``non-planar'' in the context of rods embedded in three dimensions rather than two.}
The relative magnitudes and sign of these quantities identify a mother curve from which a particular \emph{elastica} may be cut in order to satisfy end moment conditions.
The simplicity of the system is apparent when described in terms of conserved quantities, with angles or curvatures as dependent variables.  Connections between forces, moments, and slopes are also relatively clean in this general viewpoint.  Cumbersome expressions arise only when conditions on total length and boundary positions are imposed on particular solutions.  In practice, these conditions are satisfied numerically.  We do not dwell on these issues here, nor do we retread well known but rarely used impractical expressions for the shapes involving elliptic functions.

There are three widely used approaches to the planar \emph{elastica}.  Classical continuum mechanics employs Cosserat directors and begins with balances of linear and angular momentum \cite{Antman05}, which can be combined into a single vector equation.  The other two are variational treatments leading either to a single scalar pendulum equation for the tangential angle, or a single vector equation from which one can obtain a scalar equation for the curvature as the position of a particle in either a single- or double-well potential.  
These scalar shape equations represent the normal force balance on the body.  
In this note, we will relate the natural quantities in the vector and shape equations.
In variational treatments, the magnitude of the stress and the only component of the material stress appear as first integrals.  Though not typically expressed in these terms, in the pendulum approach it is customary to fix the stress and sweep the material stress to obtain a phase portrait.  However, the other approach clearly indicates that changes in material stress lead to a pitchfork bifurcation between single- and double-well potentials.  We show that this has physical meaning.
Overall, it can be said that the pendulum description fails to capture information about tangential balance that can at times provide a more direct solution of physical problems involving elastic curves.
Curiously, angular momentum balance and its associated conserved torque do not play any necessary role in this simple system, although the pendulum equation can be interpreted as a moment balance rather than a normal force balance.

We begin with a variational derivation of the vector equation and boundary conditions in Section \ref{Sec:vector_formulation}, including a discussion of conserved quantities and classical rod theories in Section \ref{Sec:symmetries} and of the shape equation for the curvature in Section \ref{Sec:shape}.  This shape equation can be derived from the conserved stress and material stress, or by manipulation of the Euler-Lagrange equation.  
 The variational scalar pendulum description is briefly reviewed in Section \ref{Sec:pendulum_description}. In Section \ref{Sec:classification}, we propose a classification scheme based on conserved quantities, and discuss phase portrait representations of both scalar shape equations.
In Section \ref{Sec:discussion}, we present several examples involving physical boundary conditions on forces, moments, slopes, and the length of the elastic curves.

\section{Vector approaches}\label{Sec:vector_formulation}

\begin{figure}[h!]
	\includegraphics[width=12cm]{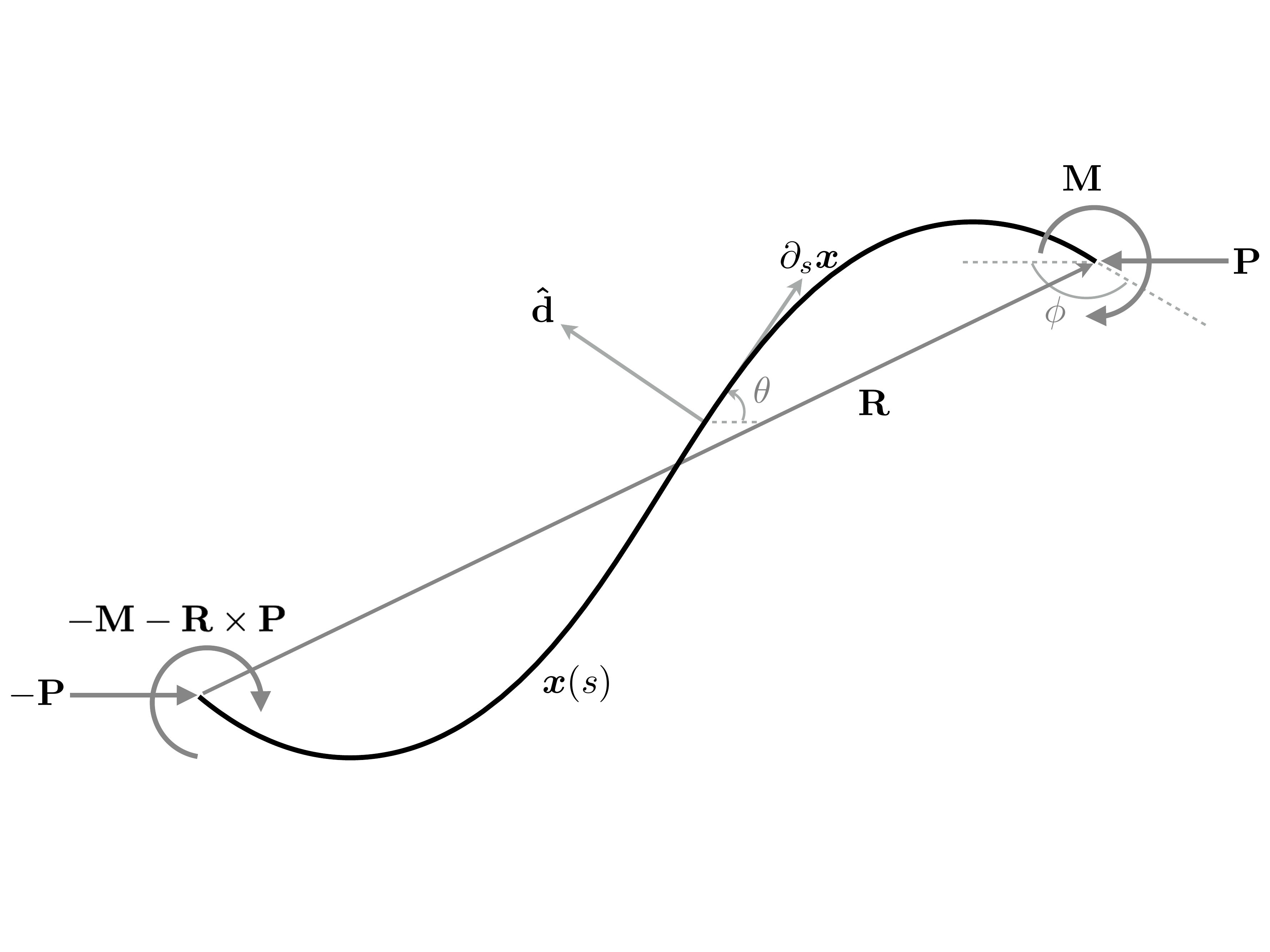}
	\vspace{-1.5cm}
	\caption{An elastic curve $\bX(s)$ with end-to-end vector $\bR$ and local frame of unit tangent $\partial_s\bX$ and unit normal $\uvc{d}$.  Also shown are a terminal force $\bP$ and moment $\bM$, the corresponding force and moment at the other end, the local tangential angle $\theta$ between $\partial_s\bX$ and $\bP$, and the angle $\phi$ between a terminal tangent and $\bP$.  }
	\label{Fig:elastica}
\end{figure}

The planar Euler \emph{elastica} is a one-dimensional idealized model of a thin, uniform, inextensible, unshearable rod or tape without rest curvature.  This body is represented by the position (embedding) vector of a planar curve $\bX(s)$, where $s \in [0,l] $ is both a material coordinate and the arc length.  An example is shown in Figure \ref{Fig:elastica}, decorated with various quantities to be introduced shortly.  The elastic energy, given by the square of the curvature of $\bX$, is expressible in several different ways.  Perhaps the simplest form for the static Lagrangian is entirely in terms of derivatives of position,
\begin{align}
L = \int_0^l \!\!ds\, \mathcal{L}(\bX) = \int_0^l \!\!ds \left[\tfrac{1}{2}B\partial_s^2\bX\cdot\partial_s^2\bX + \tfrac{1}{2}\sigma\left(\partial_s\bX\cdot\partial_s\bX - 1\right)\right]\, ,\label{lagrangian}
\end{align}
where the first term is the bending energy with uniform modulus $B$ (here with units of force), and the second is a quadratic constraint on the local length, enforced by a multiplier $\sigma(s)$.
While $\sigma$ contributes to the tension in the body, the full tension contains contributions from the curvature as well.\footnote{Similar multipliers can be found in Burchard and Thomas \cite{BurchardThomas03}, Singer \cite{Singer08}, Tornberg and Shelley \cite{TornbergShelley04}, who do not employ a variational derivation, and Guven and V{\'{a}}zquez-Montejo \cite{GuvenVasquezMontejo2012}, who use multiple redundant multipliers.  The multipliers in \cite{BurchardThomas03} and \cite{TornbergShelley04} are misidentified as the tension.  For seemingly similar but qualitatively different multipliers, see the following footnote.}

Under a small shift in the position vector $\bX\rightarrow\bX+\delta\bX$, the first order variation in the Lagrangian is
\begin{align}
\delta L^{(1)} = \left. \left[B\partial_s^2\bX\cdot\partial_s\delta\bX + (\sigma\partial_s\bX - B\partial_s^3\bX)\cdot\delta\bX\right] \right|_0^l + \int_0^l \!\!ds \left[B\partial_s^4\bX - \partial_s(\sigma\partial_s\bX)\right]\cdot\delta\bX\, .\label{lagrangian_shift}
\end{align}
Setting $\delta L^{(1)}=0$ for arbitrary $\delta\bX$ and $\partial_s\delta\bX$ yields the bulk field equation
\begin{align}
B\partial_s^4\bX - \partial_s(\sigma\partial_s\bX) = 0 
\, , \label{force_bulk}
\end{align}
and boundary conditions of the form
\begin{align}
\sigma\partial_s\bX - B\partial_s^3\bX &= \bP \quad\text{at}\; s=l \, ,\label{force_boundary}\\
B\partial_s^2\bX &= \bQ \quad\text{at}\; s=l \, ,\label{moment_boundary}
\end{align}
where $\bP$ and $\bQ$ are per-area sources of force and torque at one boundary.  Boundary conditions at the other end can be written analogously.  While an equal but opposite force must balance $\bP$, the torques at the two boundaries can differ.
 The force $\bP$ could also be included in the Lagrangian using a term $\left.-\bP\cdot\bX\, \right|_0^l = -\bP\cdot\bR$, as is done by several authors \cite{Tsuru1986, SteigmannFaulkner1993, KehrbaumMaddocks1997, GuvenVasquezMontejo2012}.  The quantities in the boundary condition \eqref{moment_boundary} live in the plane, but it is customary to write things in terms of the out-of-plane moment $\bM = \partial_s\bX\times\bQ$ applied to the boundary.
 Applying the cross product with the tangent to \eqref{moment_boundary}, we may write an equivalent boundary condition \begin{align}
\partial_s\bX\times B\partial_s^2\bX &= \bM \quad\text{at}\; s=l \, .\label{moment_boundary2}
\end{align}
Should we wish to obtain this boundary condition directly, one option is to write the bending energy in \eqref{lagrangian} as $\tfrac{1}{2}B\bOmega\cdot\bOmega$, where $\bOmega=\partial_s\bX\times\partial_s^2\bX$ is the Darboux vector for a planar curve. Varying the position vector leads to the same variation of the Lagrangian \eqref{lagrangian_shift}, except that the first boundary term becomes $(\partial_s\bX\times B\partial_s^2\bX)\cdot(\partial_s\bX\times\partial_s\delta\bX)$. We may then treat $\delta\bX$ and $\partial_s\bX\times\partial_s\delta\bX$ as our two arbitrary variations yielding two boundary conditions. 
There are many ways to express the square of the curvature in this simple system.  Another is to square the  wryness, defined as ${\bm d}\times\nabla {\bm d}\,$, where $\bm d$ is the director, here identified with the unit normal.

We may connect our equations and boundary conditions with classical rod theory \cite{Antman05} by identifying the (per-area) contact force $\bn(s)$ and contact moment $\bmo(s)$ as
\begin{align}
\bn&\equiv\sigma\partial_s\bX - B\partial_s^3\bX\, ,\label{contact_force}\\
\bmo&\equiv\partial_s\bX\times B\partial_s^2\bX\, .\label{contact_moment}
\end{align}
The contact force\footnote{The tension $\bn\cdot\partial_s\bX$ can be associated with the multiplier $T$ appearing in Nordgren \cite{Nordgren1974} and Shelley and Ueda \cite{ShelleyUeda00}, who do not employ variational derivations, and Audoly \cite{Audoly2016}, whose multiplier is different than our $\sigma$ despite appearing similarly in a Lagrangian.  The difference with Audoly arises due to his splitting of the variation in terms of an angular variation $\delta\theta$ on the boundary, rather than $\partial_s\delta\bX$ as done here.  A similar issue arises in plate and shell theories \cite{Steigmann12}.}
 can be identified as the term conjugate to the variation $\delta\bX$ in the boundary term in \eqref{lagrangian_shift}, while the contact moment is conjugate to $\partial_s\bX\times\partial_s\delta\bX$ in the Darboux variation described above.

\subsection{Symmetries}\label{Sec:symmetries}

The Lagrangian \eqref{lagrangian} does not depend on position and orientation in space or on the choice of origin for the coordinate $s$.  These translational and rotational symmetries of space and material symmetry of the body imply the existence of conserved quantities.

For a shift in the embedding vector $\bX$ corresponding to a symmetry of the Euclidean plane, the current can be read off from the boundary term in \eqref{lagrangian_shift}.  Translations correspond to a uniform $\delta\bX$, so the contact force is conserved along the body,
\begin{align}
\sigma\partial_s\bX-B\partial_s^3\bX = \bn= \bP\, .\label{P}
\end{align}
We have identified the conserved stress with $\bP$ using the force boundary condition \eqref{force_boundary}.  That $\bn$ is conserved may also be seen by integrating the bulk equation \eqref{force_bulk}, which in terms of the contact force is simply
\begin{align}
-\partial_s\bn=0 \,  . \label{force_balance}
\end{align}
Rotations correspond to a shift in the position vector of the form $\delta\bX =\bC\times\bX$, where $\bC$ is a constant vector.  The boundary term yields the conserved torque,
\begin{align}
\partial_s\bX\times(B\partial_s^2\bX)+\bX\times(\sigma\partial_s\bX-B\partial_s^3\bX) = 
\bmo + \bX\times\bn = \bJ\, .\label{J}
\end{align}
Unlike the contact force $\bn$, the contact moment $\bmo$ is not conserved.  
Differentiating \eqref{J} with respect to $s$, and using \eqref{force_balance}, we obtain the bulk moment balance
\begin{align}
\partial_s\bmo+\partial_s\bX\times\bn = 0\, .\label{moment_balance}
\end{align}
For our \emph{elastica}, it can be seen from \eqref{contact_force} and \eqref{contact_moment} that $\partial_s\bX\times\bn = -\partial_s\bX\times B\partial_s^3\bX$ and $\partial_s\bmo = \partial_s\bX\times B\partial_s^3\bX$, so that the moment balance  \eqref{moment_balance} is identically satisfied, and does not provide any additional information beyond the force balance.  The torque $\bJ$ can always be eliminated by a shift in the origin $\bX\rightarrow\bX + \tfrac{\bn}{\bn\cdot\bn} \times\bJ$ by noting that $\bmo\cdot\bn=\bJ\cdot\bn=\bm 0$ in our system \cite{KehrbaumMaddocks1997}.

The third symmetry corresponds to a shift in the independent variable $s$ by a uniform $\delta s$.  This transformation is associated with the uniformity of the body, and has nothing to do with the embedding space.
We denote the transformed material label as $\shift{s} \equiv s+\delta s$.  The vector $\shift{\bX}(\shift{s})$ corresponds to the same material point as $\bX(s)$ did in the original coordinates, while $\shiftvar{\delta}\bX \equiv \shift{\bX}(s) - \bX(s)$ is the shift in the position vector at the same material label due to the shift $\delta s$ (\emph{i.e.} $\shift{\bX}(s) \equiv \left. \shift{\bX}(\shift{s}) \right|_{\shift{s}=s}$ is at the same value of $s$, not the same material point).  Note that $\shiftvar{\delta}$ and $\partial_s$ commute.
 For a detailed presentation of action principles, including variations in the independent coordinates, the reader is referred to \cite{NoetherTavel1971, Hill1951}.
Under a small uniform shift in the material coordinate $s\rightarrow \shift{s}$, the variation in the Lagrangian is
\begin{align}
L + \delta L &= \int_{s_1+\delta s}^{s_2 +\delta s} \!\!ds\, \mathcal{L}(\bX+\shiftvar{\delta}\bX)\\
&= L + \left.\mathcal{L}(\bX)\delta s \right|_{s_1}^{s_2} + \int_{s_1}^{s_2}\!\!ds \left[B\partial_s^2\bX\cdot\partial_s^2\shiftvar{\delta}\bX+\sigma\partial_s\bX\cdot\partial_s\shiftvar{\delta}\bX\right] + \text{higher order terms.}
\end{align}
 Integrating by parts and extremizing by setting the first order terms to zero,
\begin{align}
0 = \left.\left[\mathcal{L} \delta s + B\partial_s^2\bX\cdot\partial_s\shiftvar{\delta}\bX + (\sigma\partial_s\bX - B\partial_s^3\bX)\cdot\shiftvar{\delta}\bX\right] \right|_{s_1}^{s_2} + \int_{s_1}^{s_2}\!\!ds \left[B\partial_s^4\bX - \partial_s(\sigma\partial_s\bX)\right]\cdot\shiftvar{\delta}\bX \, .
\end{align}
For a uniform material translation $\delta s$, the current may be associated with the boundary term after using the condition $\shift{\bX}(\shift{s}) = \bX(s)$ to express the variations in the position and the tangent as $\shiftvar{\delta}\bX=-\partial_s\bX\,\delta s$ and $\partial_s\shiftvar{\delta}\bX=- \partial_s^2\bX\,\delta s$.
 Using the constitutive relations (\ref{contact_force}-\ref{contact_moment}) for the contact force and moment, we obtain the only component of the conserved material force (per area),
\begin{align}
-\tfrac{1}{2}B\partial_s^2\bX\cdot\partial_s^2\bX +\bmo\cdot\bOmega + \bn\cdot\partial_s\bX =\tfrac{\bmo\cdot\bmo}{2B} + \bn\cdot\partial_s\bX= c \, . \label{c}
\end{align}
We have chosen to reverse the sign of the boundary term to align with the definition of stress in a more general setting that includes inertia, such that the time derivative of the momentum equals the divergence of stress, and the time derivative of the projection of the momentum onto the tangents equals the divergence of material stress--- compare with equation (8) of Broer \cite{Broer70} and equation (3.2) of Maddocks and Dichmann \cite{MaddocksDichmann1994}.
O'Reilly \cite{OReilly07, OReilly2017} defines the equivalent quantity with the opposite sign, in keeping with the conventional form for the Eshelby tensor.  
More general forms of $c$ may be found in the context of static \cite{Ericksen70, AntmanJordan75, NizetteGoriely99, VanderHeijdenThompson00} and dynamic \cite{MaddocksDichmann1994} spatial hyperelastic rods.
Another way to obtain this quantity is to integrate the tangential component of the force balance \eqref{force_balance}. 
Integrating $\partial_s\bn\cdot\partial_s\bX$ by parts and anticipating the identity of the constant, we can write
$c = \bn\cdot\partial_s\bX - \int \!\!ds\, \bn\cdot\partial_s^2\bX$,
which, upon inserting (\ref{contact_force}-\ref{contact_moment}), simplifies to $c=\bn\cdot\partial_s\bX + \tfrac{\bmo\cdot\bmo}{2B}$.  

Curiously, the conserved material stress can appear directly as a multiplier in an approach \cite{GoldsteinLanger95, Capovilla02} that employs a fully covariant description of a \emph{geometric} energy, $\int\!\! \sqrt{g}\,d\alpha \left(\tfrac{1}{2} \nabla^2\bX\cdot\nabla^2\bX + c \right)$.  In this formalism, common in some areas of physics, the volume form $\sqrt{g}\,d\alpha$ varies, and the energy density is to be interpreted as per present volume rather than per mass, a distinction that only matters when the rod is extensible.

\subsection{Shape equation}\label{Sec:shape}

The shape of a planar curve may be reconstructed from the tangential angle $\theta(s)$ or its derivative, the curvature $\kappa = \partial_s\theta$.
For simplicity, we make use of an adapted orthonormal frame of tangent vector $\partial_s\bX$ and normal $\uvc{d}$, such that $\partial_s^2\bX=\kappa\uvc{d}$ and $\partial_s\uvc{d} = -\kappa\partial_s\bX$. 
Using the definitions of the conserved quantities \eqref{P} and \eqref{c} along with the constitutive relations (\ref{contact_force}-\ref{contact_moment}), we may write
\begin{align}
(\sigma + B\kappa^2)\,\partial_s\bX - B\partial_s\kappa\, \uvc{d} &= \bP \, ,\label{P_frenet}\\
\sigma + \tfrac{3}{2} B \kappa^2 &= c \, . \label{c_frenet}
\end{align}
Eliminating the multiplier $\sigma$ and squaring \eqref{P_frenet} leads to the shape equation
\begin{align}
(B\partial_s\kappa)^2 + (\tfrac{1}{2}B\kappa^2 - c)^2 = P^2\, ,\label{shape_equation}
\end{align}
where $P^2 = \bP\cdot\bP$.  This is a restricted form of a more general equation derived by Lu and Perkins \cite{LuPerkins94} for symmetric spatial Kirchhoff rods.

Alternatively, we can derive the shape equation \eqref{shape_equation} by projecting the bulk Euler-Lagrange equation \eqref{force_bulk} onto the frame to obtain two equations,
\begin{align}
B\left(\partial_s^2\kappa - \kappa^3\right) - \sigma\kappa &= 0 \, ,\label{bulk_normal}\\
-3B\kappa\partial_s\kappa - \partial_s\sigma &=0 \, . \label{bulk_tangential}
\end{align}
Integrating \eqref{bulk_tangential} and switching signs gives equation \eqref{c_frenet}, and then inserting for $\sigma$ in \eqref{bulk_normal} gives
\begin{align}
B\left(\partial_s^2\kappa + \tfrac{1}{2}\kappa^3\right) -c\kappa = 0 \, . \label{shape_secondorder}
\end{align}
Multiplying \eqref{shape_secondorder} by $2B\partial_s\kappa$, integrating, and calling the constant of integration $P^2-c^2$ recovers the shape equation \eqref{shape_equation}.
The associated scalar boundary conditions derived from \eqref{force_boundary} and \eqref{moment_boundary2} are 
\begin{align}
	\sigma + B\kappa^2 = \bP\cdot\partial_s\bX  \quad\text{at}\; s=l \, , \label{bc_tangential}\\
	-B\partial_s\kappa = \bP\cdot\uvc{d}  \quad\text{at}\; s=l \, , \label{bc_normal}\\
	B\kappa = M  \quad\text{at}\; s=l \, , \label{bc_moment}
\end{align}
where $M$ is the only component of the moment at one boundary.  Note that either \eqref{c}, or \eqref{c_frenet} along with the boundary conditions \eqref{bc_tangential} and \eqref{bc_moment}, lead to the relation
\begin{align}
c = \tfrac{M^2}{2B}+P\cos\phi\, ,\label{c_pendulum}
\end{align}
where $\phi$ is the angle between $\bP$ and the boundary tangent $\partial_s\bX(l)$.  
Hence, the material stress can be determined easily from the magnitude and direction of the end force, and the single component of the end moment.

The curvature $\kappa$ uniquely defines a planar curve up to rigid translations and rotations.  There are the two parameters $P$ and $c$ and one additional degree of freedom corresponding to the final integration of the shape equation.
However, the physical conditions one might impose on an elastic curve are related in a complicated way to these degrees of freedom.  
One might, for example, specify the ratio of the body's length to its end to end distance, along with the two angles made by the end tangents $\partial_s\bX$ and the end to end vector $\bR$.
Or one might specify another quantity such as the moment $M$, or the angle $\phi$ between the force and the end tangent.  Note that one can specify either the end position or $\bP$, and one can specify either the end tangent or $\bM$.
 The curve may be reconstructed using $\partial_s\theta = \kappa$ and $\partial_s\bX =  \left(\cos\theta, \sin\theta \right)$.  We also have $\bR = \bX(l)-\bX(0)$ and $l = s(l)-s(0)= \int_0^l\! ds \sqrt{\partial_s\bX\cdot\partial_s\bX} = \int_{\kappa(0)}^{\kappa(l)}  d\tilde{\kappa}/\partial_s\tilde{\kappa}$.
The significant complications that arise in applications of an \emph{elastica} model are often due to specifications of length and position.

In Section \ref{Sec:classification}, we will detail how the two parameters in the shape equation define shapes of infinite length, and how the moment boundary condition chooses the end point of any shape cut from such a mother curve.

\section{Scalar approach}\label{Sec:pendulum_description}

Before presenting our classification and examples, we briefly revisit the most well known description of the planar \emph{elastica}, which gives rise to a pendulum equation.
We may write the Lagrangian in terms of the tangential angle, rather than the position vector,
\begin{align}
L(\theta) = \int_0^l \!\!ds \left[\tfrac{1}{2}B(\partial_s\theta)^2\right] + P \left[\int_0^l \!\!ds\left(\cos\theta\right) - R_P \right]\, , \label{lagrangian_pendulum}
\end{align}
where $P$ now appears as a global constraint on the distance $R_P$ between the ends in the direction of the force.  There is no constraint term corresponding to the direction orthogonal to the force.
We have measured $\theta$ with respect to the direction of force, a quantity not always experimentally known, because it directly leads to the simplest form of the Euler-Lagrange equations.  

Extremizing the Lagrangian with respect to shifts in the angle $\theta\rightarrow\theta+\delta\theta$ leads to the bulk field equation
\begin{align}
B\partial_s^2\theta+ P\sin\theta &= 0  \, ,\label{bulk_pendulum_1}
\end{align}
and boundary conditions of the form
\begin{align}
B\partial_s\theta &=M\quad\text{at}\; s=l \, ,\label{moment_boundary_pendulum}
\end{align}
the latter equivalent to \eqref{bc_moment}.  The moment term could be included in the Lagrangian using a term $-M\theta(l)$.  
The pendulum equation \eqref{bulk_pendulum_1} is equivalent to the projection of the conserved stress \eqref{P} onto the unit normal $\uvc{d}$, as well as the only component of the moment balance \eqref{moment_balance}.  Upon integration, we recover the conserved material stress,
\begin{align}
\tfrac{1}{2}B(\partial_s\theta)^2 - P\cos\theta = c\, .\label{pendulum_shape}
\end{align}
This shape equation is equivalent to the definition \eqref{c}.  With the identification of the curvature with the end moment \eqref{moment_boundary_pendulum}, we recover \eqref{c_pendulum}.  The pendulum equation \eqref{pendulum_shape} is a restricted form of a more general oscillator equation derived by van der Heijden and Thompson \cite{VanderHeijdenThompson00} for symmetric spatial Kirchhoff rods.

It is not immediately obvious that the shape equations \eqref{shape_equation} and \eqref{pendulum_shape} represent the same curves.  
However, using \eqref{bulk_pendulum_1} and \eqref{pendulum_shape} to square and add the trigonometric terms, and recalling that $\kappa = \partial_s\theta$, we recover \eqref{shape_equation}.
We will return to the comparison between curvature and angle shape equations in Section \ref{Sec:classification}, in the context of phase portrait representations of these equations.

\section{Classification in terms of stress and material stress}\label{Sec:classification}

\begin{figure}[h!]
	\begin{subfigure}[t]{16cm}
		\includegraphics[width=\textwidth]{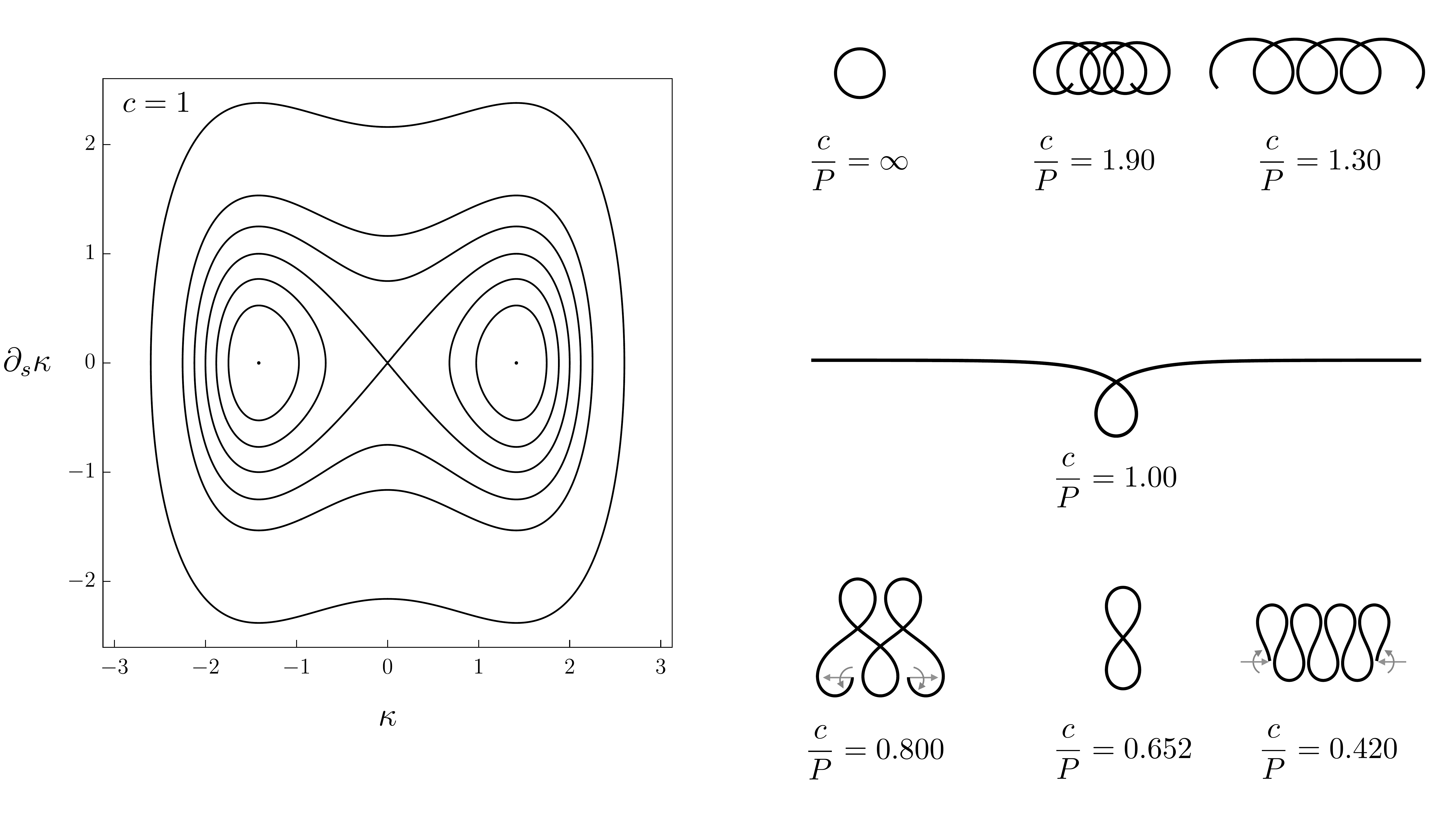}
		\label{Fig:phaseplotspositivec}
	\end{subfigure}\\
			\vspace{-0.5cm}
	\begin{subfigure}[t]{16cm}
		\includegraphics[width=\textwidth]{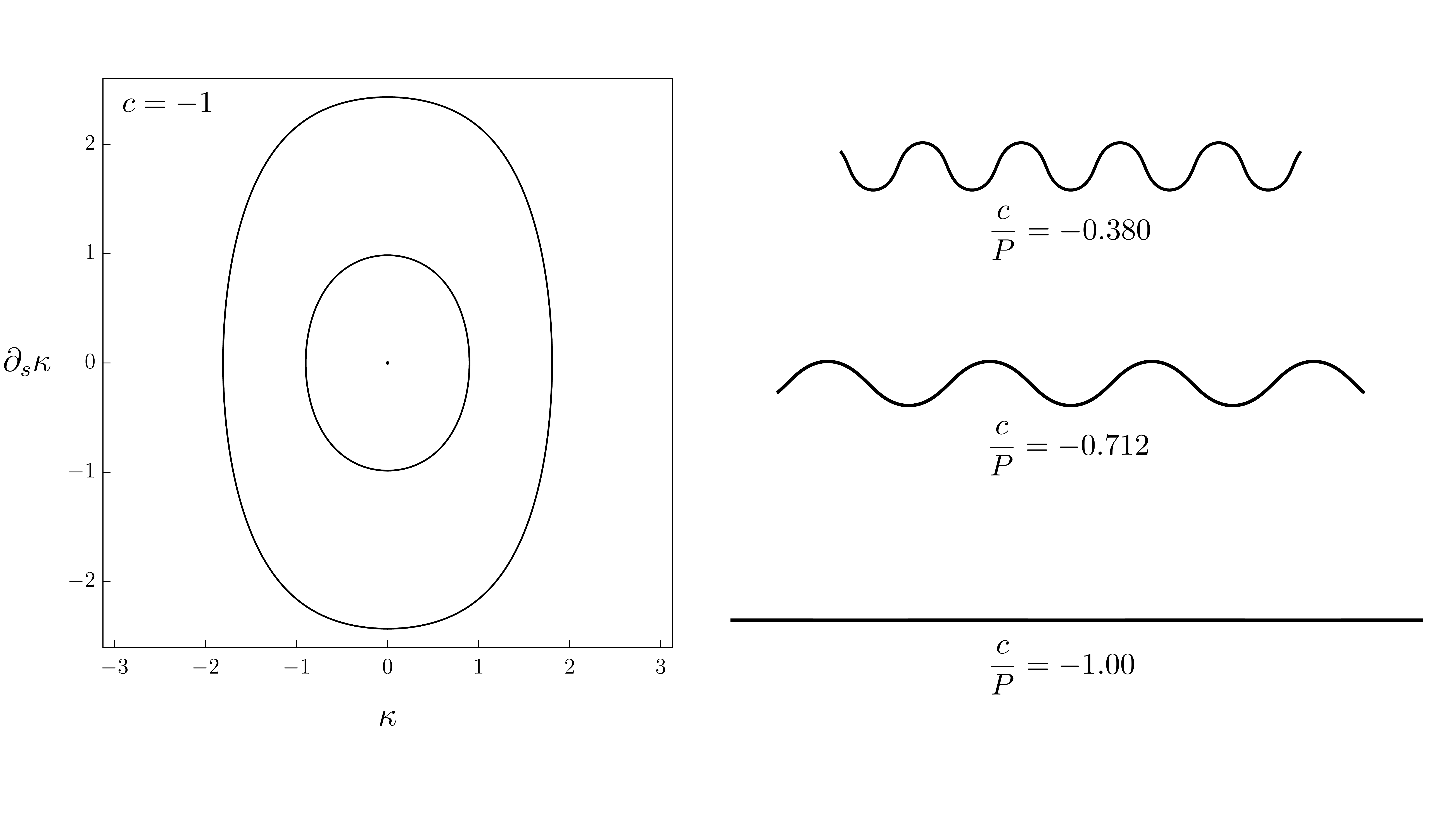}
		\label{Fig:phaseplotsnegativec}
	\end{subfigure}\\
		\vspace{-1cm}
	\caption{Phase portraits in $\kappa$-$\partial_s\kappa$ space and configurations of \emph{elastica} curves, with larger contours corresponding to larger values of stress $P$.  Arrows indicate the direction of forces for configurations in the sequence on either side of the figure-eight solution.  Moments can be inferred from the curvature of the end points.  At top, the material stress $c=1$ and the potential has two wells at $(\pm\sqrt{2},0)$.   Inside the $\tfrac{c}{P}=1$ separatrix, the curves are non-inflectional.   At bottom, $c=-1$ and the potential has one well at $(0,0)$.  These solutions are purely compressive. The length of each non-periodic configuration is 20.}
	\label{Fig:phase_space_kappa}
\end{figure}

The shapes of \emph{elastica} curves are governed by the two parameters in the shape equation for the curvature \eqref{shape_equation}, which under the rescalings $s\rightarrow s/L$, $\kappa\rightarrow\kappa L$, $c\rightarrow c L^2/B$, $P\rightarrow P L^2/B$, and $M\rightarrow M L/B$ ($L$ an arbitrary length scale) 
 takes the nondimensional form 
\begin{align}
(\partial_s\kappa)^2 + (\tfrac{1}{2}\kappa^2- c)^2  = P^2\, . \label{shape_non_dimensional}
\end{align}
This equation describes the motion of a particle with position $\kappa$ and energy $\tfrac{1}{2}P^2$ in a potential $\tfrac{1}{2}\left(\tfrac{1}{2}\kappa^2 - c\right)^2$.  This potential has a pitchfork bifurcation at $c=0$, such that for negative $c$ there is a single minimum at $\kappa=0$ and we require $P^2 > c^2$, and for positive $c$ there are two wells with minima at $\kappa = \pm\sqrt{2c}\,$, 
the particle being confined to one of these wells if $P^2 < c^2$.  Shapes are often classified as inflectional or non-inflectional \cite{Love}; if we define $P$ as a positive quantity, the magnitude of the conserved stress,  we can express the above relations as 
\begin{equation}
\begin{alignedat}{3}\label{criteria}
&\frac{c}{P}\in[-\infty,-1)\quad&&\text{no solution}\, , \\
&\frac{c}{P}\in[-1,0]\quad&&\text{inflectional single-well}\, , \\
&\frac{c}{P}\in(0,1)\quad&&\text{inflectional double-well}\, , \\
&\frac{c}{P}\in[1,\infty]\quad&&\text{non-inflectional}\, .
\end{alignedat}
\end{equation}
The division of the inflectional solutions into single-well and double-well type, depending on the sign of $c$, has a physical meaning.  Recall the definition \eqref{c} of the material stress $c$.  The squared moment term is strictly non-negative, so the tension $\bn\cdot\partial_s\bX \le c$.  When $c$ is negative, the rod must be in compression everywhere: the tangents $\partial_s\bX$ are always in opposition to $\bP$, and the corresponding shapes can be represented by a single-valued height function above any line parallel to the direction of the force. 
Shapes and phase portraits corresponding to double-well and single-well potentials are shown in Figure \eqref{Fig:phase_space_kappa}.
Other interesting landmarks include the straight line $\frac{c}{P}=-1$, the homoclinic solitary loop $\frac{c}{P}=1$, the circle $P=0$, and the figure-eight solution.
Clearly our formulation says nothing about the global property of self-intersection, and we refrain from commenting on this matter here although it is clearly of interest in some physical problems.  For example, in keeping with common experience, weaving somehow through the intersecting solutions must be a family of racket-like loops meeting at any angle.

The ratio $\tfrac{c}{P}$ can be simply related to parameters in elliptic integrals relating Cartesian coordinates on the curves \cite{Fraser1991, Levien2008}.  The connection with the
planar restriction of a classification of Kirchhoff rods by Nizette and Goriely \cite{NizetteGoriely99}, who use the roots of a polynomial of a cosine of an Euler angle that generates the solutions through another elliptic integral, appears to be more complicated.

\begin{figure}[h!].
	\includegraphics[width=12cm]{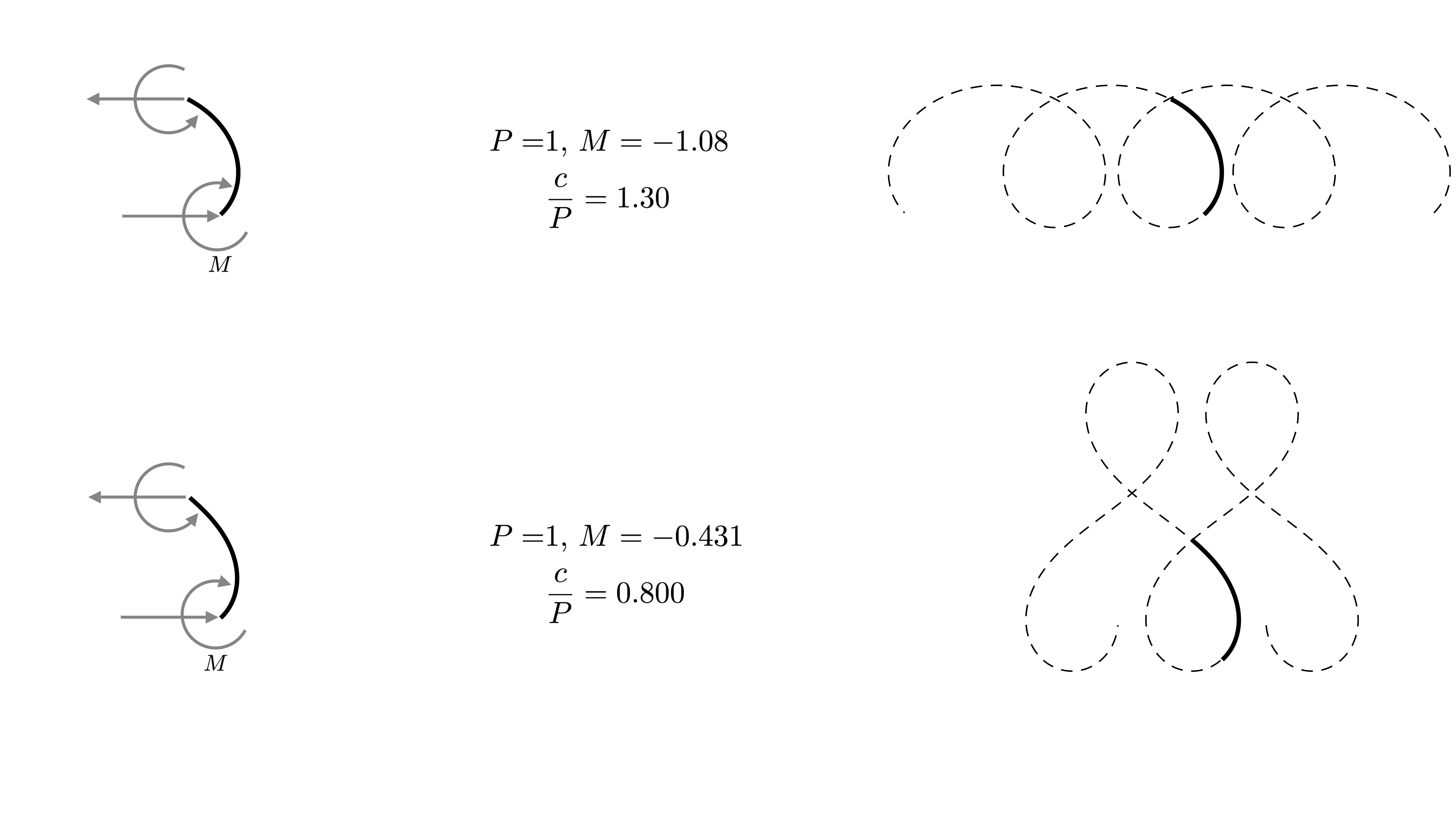}
	\vspace{-1cm}
	\caption{Two \emph{elastica} curves of length $l=1.5$ loaded with end forces and moments. From the ratio $\tfrac{c}{P}$, the top curve is identified as part of a non-inflectional shape, and the bottom curve part of an inflectional shape.}
	\label{Fig:hubbard}
\end{figure}

The physical significance of inflection points is that they are points where the moment vanishes, and so can correspond to boundaries loaded by a pure force.
 Hubbard \cite{Hubbard1980} argues that if one end of an \emph{elastica} is loaded like this, and the other end loaded with a force and a moment, then the curve can be extended to a fictitious curve loaded only with end forces.  This is simply a consequence of the fact that an \emph{elastica} with pure force loading on one end must be an inflectional \emph{elastica}, and can be extended up to the next inflection point.
Note that the absence of inflection points on a finite length of curve does not imply a non-inflectional \emph{elastica}.  Compare the two qualitatively similar curves in Figure \ref{Fig:hubbard}, cut from distinctly different mother curves, one inflectional and one not.

\begin{figure}[h!]
	\includegraphics[width=8cm]{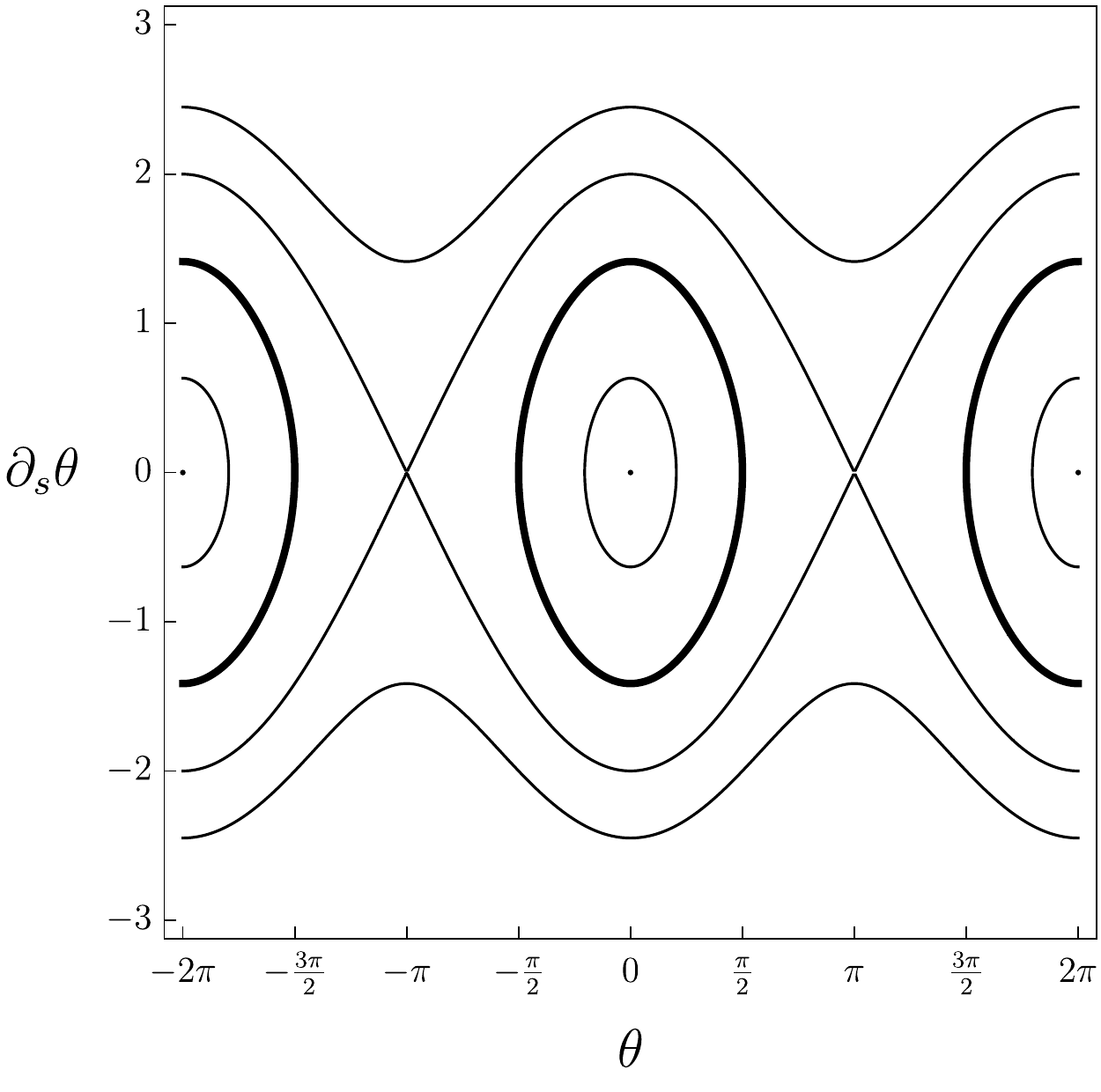}
	\caption{Phase portrait in $\theta$-$\partial_s\theta$ space with stress $P=1$ and increasing contours $\tfrac{c}{P}=(-1, -0.8, 0, 1, 2)$.  The fixed points are straight lines, and the separatrices are solitary loop solutions separating inner inflectional and outer non-inflectional solutions.  
Not apparent in this pendulum description is the fact that solutions inside the bold $\tfrac{c}{P}=0$ contour are purely compressive.}
	\label{Fig:phase_space_angle}
\end{figure}

We can also interpret a rescaled form of the pendulum equation \eqref{pendulum_shape} in terms of motion with energy $c$ in a sinusoidal potential with amplitude $P$, which again requires $\tfrac{c}{P} \ge -1$.  A phase portrait is shown in Figure \ref{Fig:phase_space_angle}.  
This is something of an inversion of the curvature description, with the roles of stress and material stress reversed, and the inner librational contours of the phase portrait now representing inflectional solutions, and the outer rotational contours non-inflectional solutions.  We may reproduce all the qualitative types of configurations shown in Figure \ref{Fig:phase_space_kappa} by varying the material stress $c$ at some fixed stress $P$. 
Inside the $\tfrac{c}{P}=0$ contour, which has the property that the tangential angle $\theta$ subtends exactly $\pi$ radians, the curves are purely compressive.  However, this fact is not apparent from the pendulum description, which does not contain any information about tangential forces and does not undergo a bifurcation.
Although the curvature description is also in the form of a scalar equation representing normal force balance, it is derived using the tangential force balance and inherits important information from it.

\section{Examples}\label{Sec:discussion}

Some examples will help relate our classification, which makes use of the somewhat esoteric conserved material force $c$, to physical quantities such as the magnitude $P$ and direction of the end force, the end moment $M$, and the length of the body.  All but the length are related by the nondimensional form of \eqref{c_pendulum},
\begin{align}
c = \tfrac{1}{2}M^2 +  P \cos\phi \, ,\label{c_2}
\end{align}
where $P$ is defined as positive and $\phi$ is the angle between $\bP$ and the boundary tangent $\partial_s\bX(l)$.  The end moment is simply the end curvature at this boundary, according to the nondimensional form of the boundary condition \eqref{bc_moment}.

\begin{figure}[h!]
	\includegraphics[width=16cm]{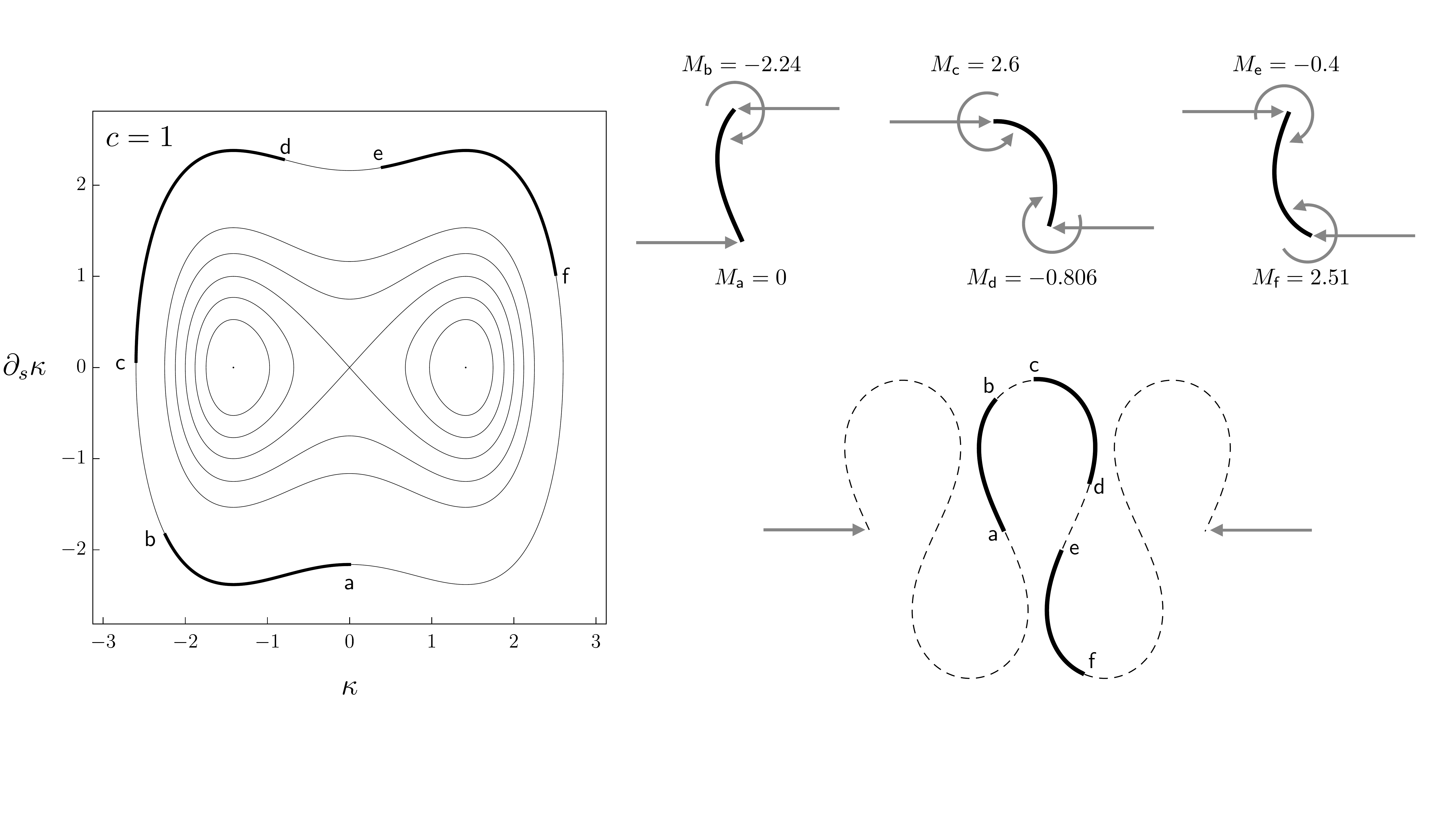}
	\vspace{-1cm}
	\caption{Orbits and configurations of a unit length \emph{elastica} corresponding to different sections of a $\tfrac{c}{P}=0.420$ mother curve.}
	\label{Fig:fixed_contour}
\end{figure}

\begin{figure}[h!]
	\includegraphics[width=16cm]{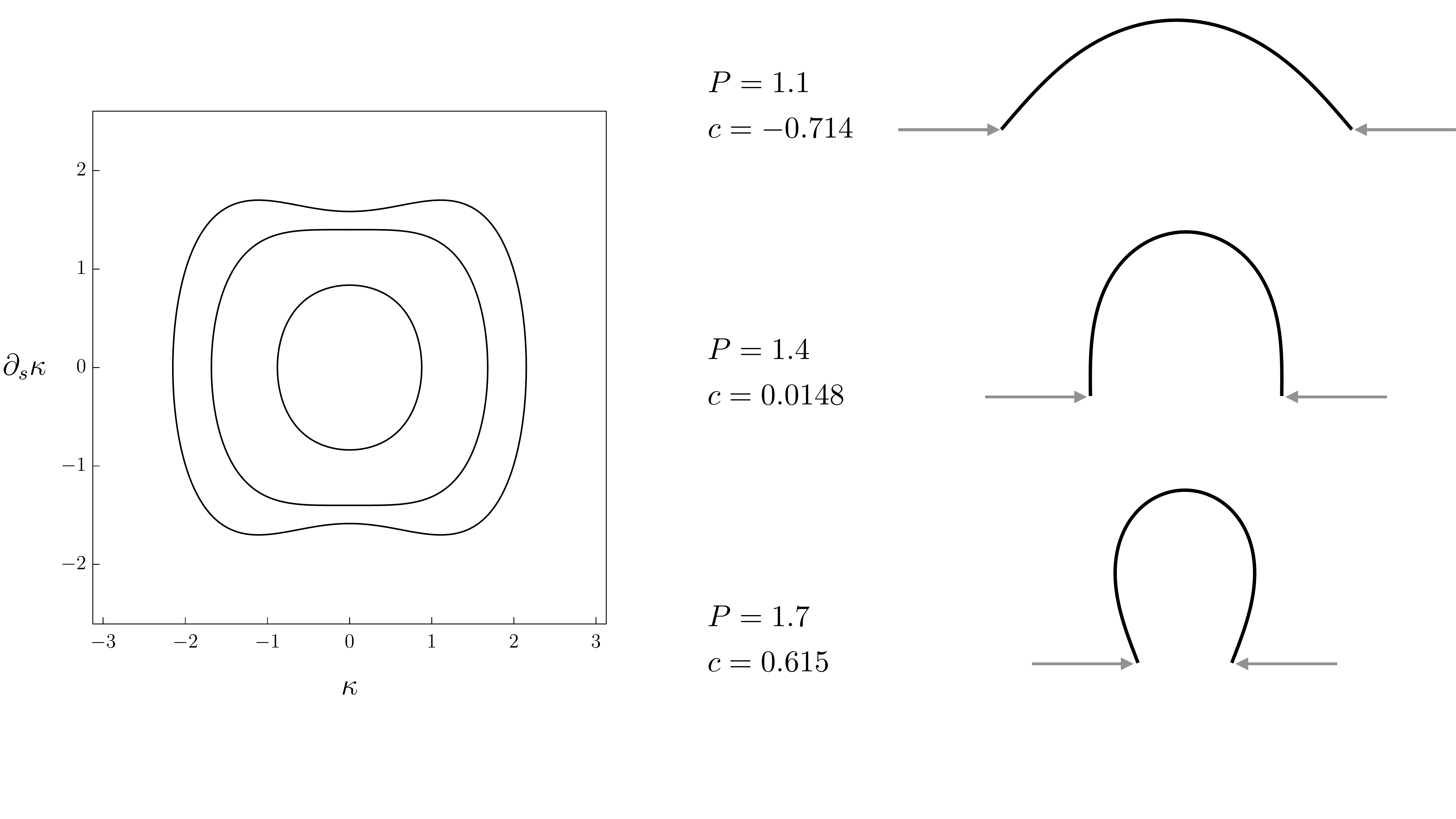}
	\vspace{-1cm}	
	\caption{Configurations of an \emph{elastica} of length $l=\pi$ under increasing end force and no end moments.  The corresponding contours, enlarging with increasing stress and material stress, are shown; the orbits begin and end on the $\partial_s\kappa$ axis and subtend half a contour.  At $c=0$, the force is orthogonal to the end tangents, and the curves transition from purely compressive to tensile near the ends and compressive in the middle.}
	\label{Fig:change_load}
\end{figure}

\begin{figure}[h!]
	\includegraphics[width=16cm]{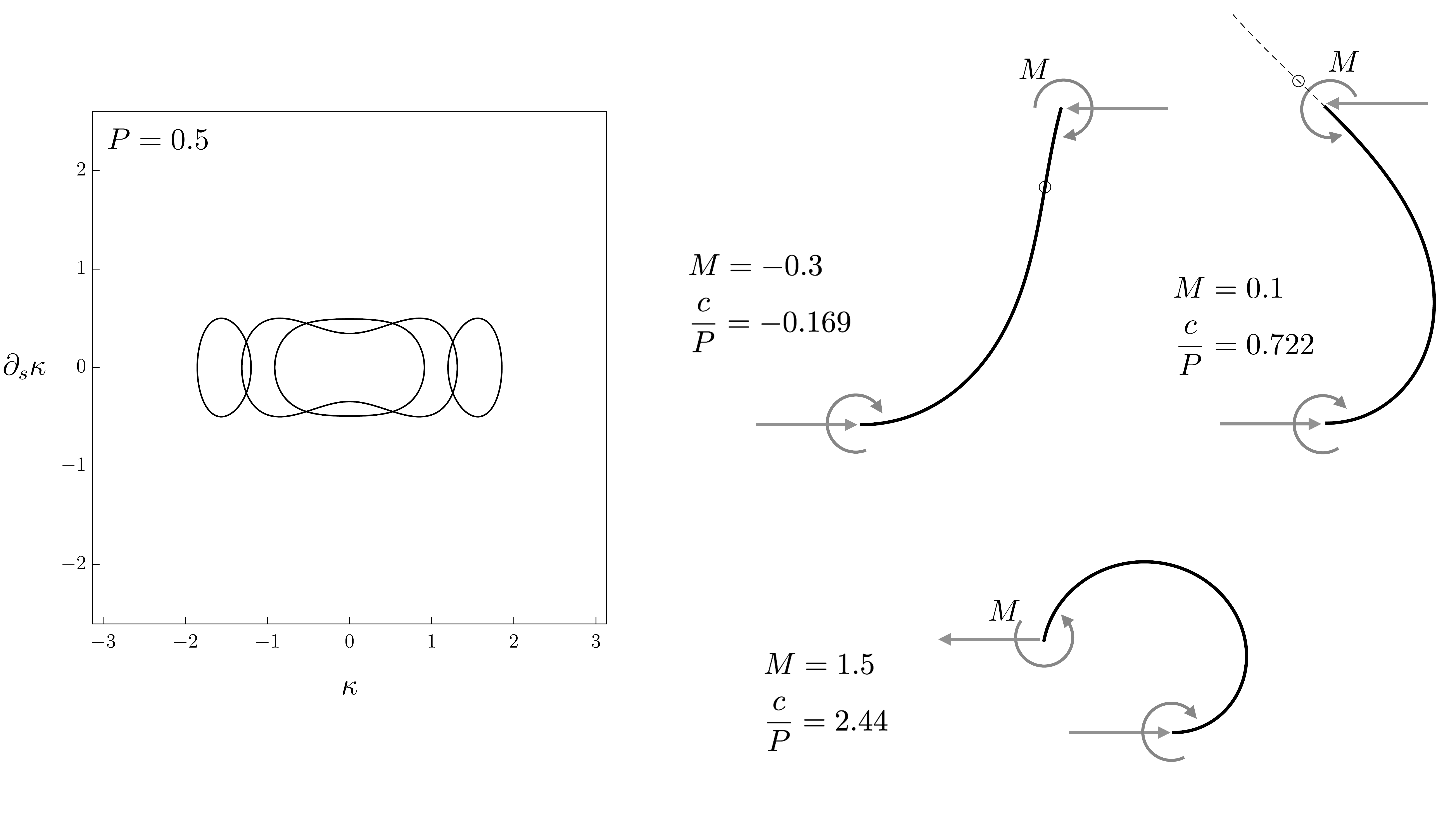}
	\vspace{-0.25cm}	
	\caption{Configurations of an \emph{elastica} of length $l=\pi$ under increasing end moment at one end, fixed slope at the other end, and fixed end forces. The corresponding contours transform from a single-well to a double-well potential.  Inflection points, real and virtual, are denoted by open circles.}
	\label{Fig:change_moment}
\end{figure}

We first observe that many related finite length curves may be cut from a single mother curve of given ratio $\tfrac{c}{P}$, by applying appropriate moments at the ends.  Figure \ref{Fig:fixed_contour} displays three such configurations of an \emph{elastica} with both $c$ and $P$ fixed.  We may slide along a contour and configuration by appropriate changes in the end moments.  In Figure \ref{Fig:change_load}, we increase the load on a segment of elastica without moments, changing the potential and energy simultaneously.
These shapes correspond to orbits subtending half a contour, beginning and ending on the $\partial_s\kappa$ axis (zero curvature).  In the absence of moments, if the stress $\bP$ is orthogonal to the end tangents, the material stress $c$ vanishes and the curves transition from purely compressive to tensile near the ends and compressive in the middle.  In Figure \ref{Fig:change_moment}, we keep the load fixed and change one end moment, while keeping the other end slope fixed.  This changes the other end moment, and the potential, while keeping the energy fixed.  The \emph{elastica} is transformed from a compressive single-well inflectional curve to a double-well inflectional curve to a non-inflectional curve living in one well of a double-well potential.  An inflection point is expelled from the body into a virtual position on a mother curve, then ceases to exist entirely. 
A related sequence without the second end moment may be found in Griner's discussion of pole-vaulting \cite{Griner84}.

\begin{figure}[h!]
	\includegraphics[width=16cm]{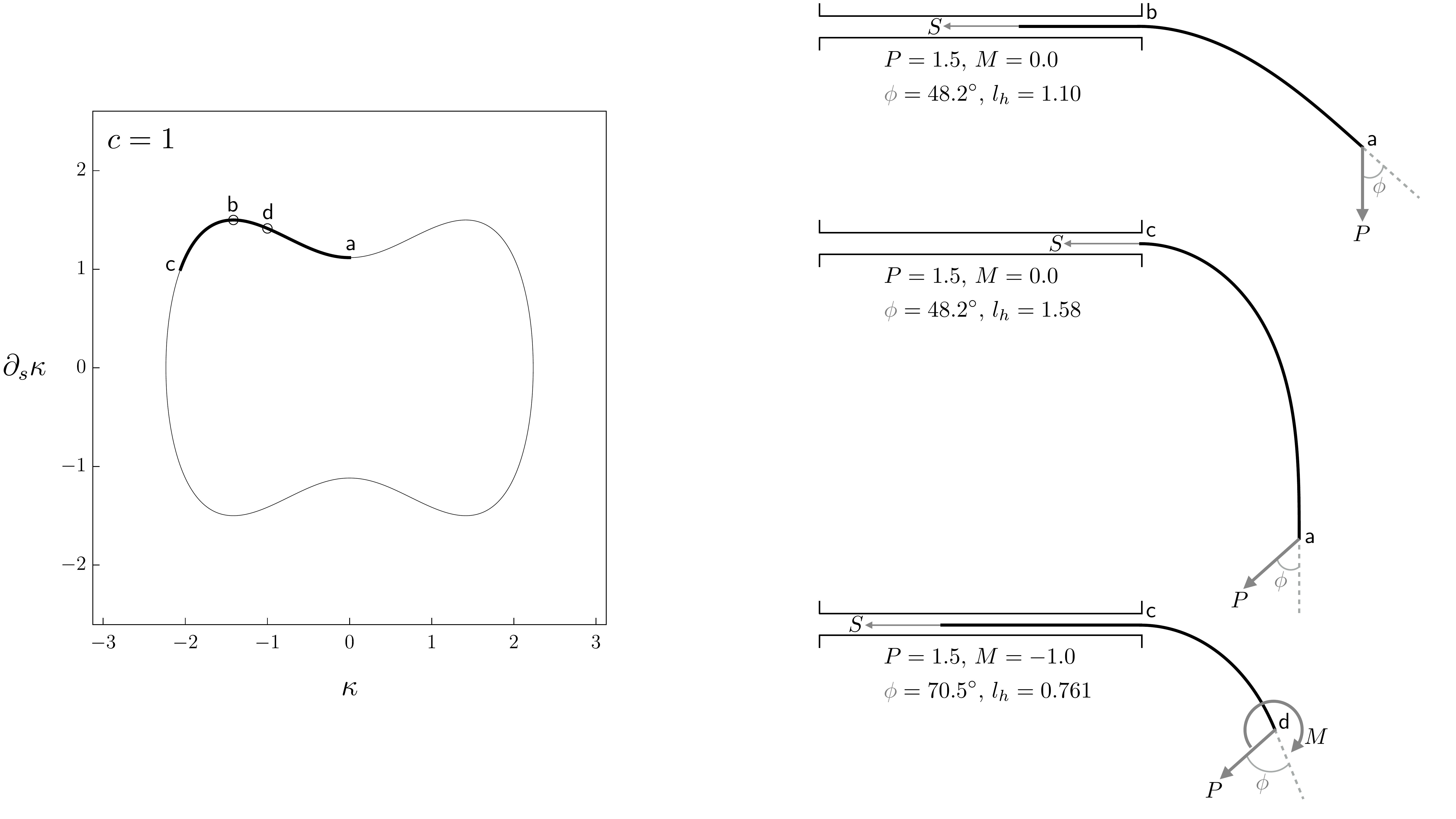}
	\caption{Configurations of a partially constrained \emph{elastica} of total length $l=1.6$, along with orbits corresponding to the freely hanging portions of lengths $l_h$ on the same $\tfrac{c}{P}=\tfrac{2}{3}$ mother contour.  The sleeve loading $S$, and thus the material stress, is the same for all configurations despite different free end loadings.  A reaction force and moment (not shown) exist at the sleeve edge.}
	\label{Fig:sleeve}
\end{figure}

Finally, we consider a system with fixed material force $c$ and variable length.  This example is inspired by O'Reilly's interpretation of a problem studied by Bigoni and co-workers \cite{OReilly2015Eshelby, Bigoni15}, featuring an elastic rod partially constrained by a straight, frictionless sleeve through which it is free to slide.  There is a jump in curvature, along with a reaction force and moment, at the sleeve edge.  A force of magnitude $S$ acts on the rod in the sleeve to hold it in equilibrium with any end force and moment.
One way to derive results on these configurations is to assume continuity of material stress at the sleeve edge \cite{OReilly2015Eshelby}.  If this is done, the quantity $S$ required to maintain the rod in equilibrium can be identified with the material force $c$ in equation \eqref{c_2}.
Note that our end moment $M$ should not be confused with the reaction moment appearing in \cite{Bigoni15}.  Figure \ref{Fig:sleeve} shows three equilibrium configurations with the same force in the sleeve and the same magnitude of end loading, so that they correspond to orbits on the same inflectional double-well contour.
The three orbits correspond to three different freely hanging lengths $l_h$.
The top and middle configurations correspond to loading by a follower force (same $\phi$), while the bottom configuration has an additional moment and a change in $\phi$.  The top configuration shares the same shape (rotated) as the portion of the middle configuration far from the sleeve, while the bottom configuration shares the same shape and orientation as the portion of the middle configuration near the sleeve.
It is not obvious whether one can easily direct the body along paths between these configurations, that is, slide the rod in and out of the sleeve by simple changes in the end loading.

\section{Summary}

We have revisited the classical problem of the planar Euler \emph{elastica}, compared commonly used representations, presented a classification in terms of conserved stress and material stress, and provided several physically inspired examples.
A scalar shape equation for the curvature derived from a vector representation contains physical information not present in a scalar shape equation for the tangential angle.

\section*{Acknowledgments}
We thank E. G. Virga for alerting us to the interesting features of the sleeve example, and J. H. Maddocks and O. M. O'Reilly for helpful discussions.
This work was supported by U.S. National Science Foundation grant CMMI-1462501.  This work has been available free of peer review on the arXiv since 6/2017.

\bibliographystyle{unsrt}

\begin{thebibliography}{10}

\bibitem{Oldfather33}
W.~A. Oldfather, C.~A. Ellis, and D.~M. Brown.
\newblock Leonhard {E}uler's elastic curves.
\newblock {\em Isis}, 20(1):72--160, 1933.

\bibitem{Mladenov17}
I.~M. Mladenov and M.~Hadzhilazova.
\newblock {\em The Many Faces of Elastica}.
\newblock Springer, Cham, 2017.

\bibitem{Rogula77}
D.~Rogula.
\newblock Forces in material space.
\newblock {\em Archives of Mechanics}, 29:705--713, 1977.

\bibitem{Herrmann81}
A.~G. Herrmann.
\newblock On conservation laws of continuum mechanics.
\newblock {\em International Journal of Solids and Structures}, 17:1--9, 1981.

\bibitem{Benjamin84}
T.~B. Benjamin.
\newblock Impulse, flow force and variational principles.
\newblock {\em IMA Journal of Applied Mathematics}, 32:3--68, 1984.

\bibitem{GurevichThellung90}
V.~L. Gurevich and A.~Thellung.
\newblock Quasimomentum in the theory of elasticity and its conservation.
\newblock {\em Physical Review B}, 42:7345--7349, 1990.

\bibitem{Nelson91}
D.~F. Nelson.
\newblock Momentum, pseudomomentum, and wave momentum: Toward resolving the
  {M}inkowski-{A}braham controversy.
\newblock {\em Physical Review A}, 44:3985--3996, 1991.

\bibitem{Maugin95}
G.~A. Maugin.
\newblock Material forces: Concepts and applications.
\newblock {\em Applied Mechanics Reviews}, 48:213--245, 1991.

\bibitem{KienzlerHerrmann00}
R.~Kienzler and G.~Herrmann.
\newblock {\em Mechanics in Material Space}.
\newblock Springer, Berlin, 2000.

\bibitem{Gurtin00}
M.~E. Gurtin.
\newblock {\em Configurational Forces as Basic Concepts of Continuum Physics}.
\newblock Springer, New York, 2000.

\bibitem{Yavari06}
A.~Yavari, J.~E. Marsden, and M.~Ortiz.
\newblock On spatial and material covariant balance laws in elasticity.
\newblock {\em Journal of Mathematical Physics}, 47:042903, 2006.

\bibitem{OReilly2017}
O.~M. O'Reilly.
\newblock {\em Modeling Nonlinear Problems in the Mechanics of Strings and
  Rods}.
\newblock Springer, New York, 2017.

\bibitem{Antman05}
S.~S. Antman.
\newblock {\em Nonlinear Problems of Elasticity}.
\newblock Springer, New York, 2005.

\bibitem{BurchardThomas03}
A.~Burchard and L.~E. Thomas.
\newblock On the {C}auchy problem for a dynamical {E}uler's elastica.
\newblock {\em Communications in Partial Differential Equations}, 28:271--300,
  2003.

\bibitem{Singer08}
D.~A. Singer.
\newblock Lectures on elastic curves and rods.
\newblock {\em AIP Conference Proceedings}, 1002:3--32, 2008.

\bibitem{TornbergShelley04}
A.-K. Tornberg and M.~J. Shelley.
\newblock Simulating the dynamics and interactions of flexible fibers in
  {S}tokes flows.
\newblock {\em Journal of Computational Physics}, 196:8--40, 2004.

\bibitem{GuvenVasquezMontejo2012}
J.~Guven and P.~V{\'a}zquez-Montejo.
\newblock Confinement of semiflexible polymers.
\newblock {\em Physical Review E}, 85:026603, 2012.

\bibitem{Tsuru1986}
H.~Tsuru.
\newblock Nonlinear dynamics for thin elastic rod.
\newblock {\em Journal of the Physical Society of Japan}, 55:2177--2182, 1986.

\bibitem{SteigmannFaulkner1993}
D.~J. Steigmann and M.~G. Faulkner.
\newblock Variational theory for spatial rods.
\newblock {\em Journal of Elasticity}, 33:1--26, 1993.

\bibitem{KehrbaumMaddocks1997}
S.~Kehrbaum and J.~H. Maddocks.
\newblock Elastic rods, rigid bodies, quaternions and the last quadrature.
\newblock {\em Philosophical Transactions of the Royal Society of London A},
  355:2117--2136, 1997.

\bibitem{Nordgren1974}
R.~P. Nordgren.
\newblock On computation of the motion of elastic rods.
\newblock {\em Journal of Applied Mechanics}, 41:777--780, 1974.

\bibitem{ShelleyUeda00}
M.~J. Shelley and T.~Ueda.
\newblock The {S}tokesian hydrodynamics of flexing, stretching filaments.
\newblock {\em Physica D}, 146:221--245, 2000.

\bibitem{Audoly2016}
B.~Audoly.
\newblock Introduction to the elasticity of rods.
\newblock In C.~Duprat and H.~A. Stone, editors, {\em Fluid-Structure
  Interactions in Low-Reynolds-Number Flows}, pages 1--24. The Royal Society of
  Chemistry, Cambridge, 2016.

\bibitem{Steigmann12}
D.~J. Steigmann.
\newblock Extension of {K}oiter's linear shell theory to materials exhibiting
  arbitrary symmetry.
\newblock {\em International Journal of Engineering Science}, 51:216--232,
  2012.

\bibitem{NoetherTavel1971}
E.~Noether and M.~A. Tavel.
\newblock Invariant variation problems and {N}oether's theorem.
\newblock {\em Transport Theory and Statistical Physics}, 3:183--207, 1971.

\bibitem{Hill1951}
E.~L. Hill.
\newblock Hamilton's principle and the conservation theorems of mathematical
  physics.
\newblock {\em Reviews of Modern Physics}, 23:253--260, 1951.

\bibitem{OReilly07}
O.~M. O'Reilly.
\newblock A material momentum balance law for rods.
\newblock {\em Journal of Elasticity}, 86:155--172, 2007.

\bibitem{Broer70}
L.~J.~F. Broer.
\newblock On the dynamics of strings.
\newblock {\em Journal of Engineering Mathematics}, 4:195--202, 1970.

\bibitem{MaddocksDichmann1994}
J.~H. Maddocks and D.~J. Dichmann.
\newblock Conservation laws in the dynamics of rods.
\newblock {\em Journal of Elasticity}, 34:83--96, 1994.

\bibitem{Ericksen70}
J.~L. Ericksen.
\newblock Simpler static problems in nonlinear theories of rods.
\newblock {\em International Journal of Solids and Structures}, 6:371--377,
  1970.

\bibitem{AntmanJordan75}
S.~S. Antman and K.~B. Jordan.
\newblock Qualitative aspects of the spatial deformation of non-linearly
  elastic rods.
\newblock {\em Proceedings of the Royal Society of Edinburgh}, 73A(5):85--105,
  1974/75.

\bibitem{NizetteGoriely99}
M.~Nizette and A.~Goriely.
\newblock Towards a classification of {E}uler-{K}irchhoff filaments.
\newblock {\em Journal of Mathematical Physics}, 40(6):2830--2866, 1999.

\bibitem{VanderHeijdenThompson00}
G.~H. M Van~Der Heijden and J.~M.~T. Thompson.
\newblock Helical and localised buckling in twisted rods: A unified analysis of
  the symmetric case.
\newblock {\em Nonlinear Dynamics}, 21:71--99, 2000.

\bibitem{GoldsteinLanger95}
R.~E. Goldstein and S.~A. Langer.
\newblock Nonlinear dynamics of stiff polymers.
\newblock {\em Physical Review Letters}, 75(6):1094--1097, 1995.

\bibitem{Capovilla02}
R.~Capovilla, C.~Chryssomalakos, and J.~Guven.
\newblock Hamiltonians for curves.
\newblock {\em Journal of Physics A}, 35:6571--6587, 2002.

\bibitem{LuPerkins94}
C.-L. Lu and N.~C. Perkins.
\newblock Nonlinear spatial equilibria and stability of cables under uni-axial
  torque and thrust.
\newblock {\em Journal of Applied Mechanics}, 61:879--886, 1994.

\bibitem{Love}
A.~E.~H. Love.
\newblock {\em A treatise on the mathematical theory of elasticity}.
\newblock Dover, 1944.

\bibitem{Fraser1991}
Craig~G. Fraser.
\newblock Mathematical technique and physical conception in {E}uler's
  investigation of the elastica.
\newblock {\em Centaurus}, 34:211--246, 1991.

\bibitem{Levien2008}
R.~Levien.
\newblock The elastica: a mathematical history.
\newblock {\em University of California, Berkeley, Technical Report No.
  UCB/EECS-2008-103}, 2008.

\bibitem{Hubbard1980}
M.~Hubbard.
\newblock An iterative numerical solution for the elastica with causally mixed
  inputs.
\newblock {\em Journal of Applied Mechanics}, 47:200--202, 1980.

\bibitem{Griner84}
G.M. Griner.
\newblock A parametric solution to the elastic pole-vaulting pole problem.
\newblock {\em Journal of Applied Mechanics}, 51:409--414, 1984.

\bibitem{OReilly2015Eshelby}
O.~M. O'Reilly.
\newblock Some perspectives on {E}shelby-like forces in the elastica arm scale.
\newblock {\em Proceedings of the Royal Society of London A}, 471:20140785,
  2015.

\bibitem{Bigoni15}
D.~Bigoni, F.~Dal~Corso, F.~Bosi, and D.~Misseroni.
\newblock Eshelby-like forces acting on elastic structures: {T}heoretical and
  experimental proof.
\newblock {\em Mechanics of Materials}, 80:368--374, 2015.

\end{thebibliography}

\end{document}